\documentclass[onecolumn]{IEEEtran}
\usepackage{graphicx}
\usepackage{multirow}
\usepackage{dblfloatfix}
\usepackage{subfigure}
\usepackage{caption}

\begin{document}

\title{Risk Prediction of a Multiple Sclerosis Diagnosis}
\author{
\IEEEauthorblockN{Joyce C. Ho\IEEEauthorrefmark{1}, 
Joydeep Ghosh\IEEEauthorrefmark{2},
KP Unnikrishnan\IEEEauthorrefmark{3}} \\
\IEEEauthorblockA{\IEEEauthorrefmark{1}
University of Texas at Austin, Austin, TX 78712\\
Email: joyceho@utexas.edu}\\
\IEEEauthorblockA{\IEEEauthorrefmark{2}
University of Texas at Austin, Austin, TX 78712\\
Email: ghosh@ece.utexas.edu}\\
\IEEEauthorblockA{\IEEEauthorrefmark{3}
NorthShore University HealthSystem, Evanston, IL 60201\\
Email: KUnnikrishnan@northshore.org}
}

\maketitle
\begin{abstract} 
Multiple sclerosis (MS) is a chronic autoimmune disease that affects the central nervous system. The progression and severity of MS varies by individual, but it is generally a disabling disease. Although medications have been developed to slow the disease progression and help manage symptoms, MS research has yet to result in a cure. Early diagnosis and treatment of the disease have been shown to be effective at slowing the development of disabilities. However, early MS diagnosis is difficult because symptoms are intermittent and shared with other diseases. Thus most previous works have focused on uncovering the risk factors associated with MS and predicting the progression of disease after a diagnosis rather than disease prediction. This paper investigates the use of data available in electronic medical records (EMRs) to create a risk prediction model; thereby helping clinicians perform the difficult task of diagnosing an MS patient. Our results demonstrate that even given a limited time window of patient data, one can achieve reasonable classification with an area under the receiver operating characteristic curve of 0.724. By restricting our features to common EMR components, the developed models also generalize to other healthcare systems.
\end{abstract}

\section{Introduction}
Multiple sclerosis (MS) is a chronic, progressive, and incurable autoimmune disease. Inflammation damages the myelin sheath, the protective coating of nerve cells, and causes signal disruption in the brain and spinal cord. The deterioration of nerve cells eventually becomes irreversible and leads to the development of disabilities. At least 1.3 million people worldwide are afflicted with MS with an average onset age of 29 years~\cite{Dua:2008uf}. The incidence and prevalence rates vary amongst countries but remains a global problem~\cite{Dua:2008uf}. Currently no cure exists for MS, but medications can help manage the symptoms, modify the disease course, and enhance the lifestyle of MS patients. Clinical trials have provided evidence that early diagnosis and treatment can slow the progression of MS, delaying the development of disabilities~\cite{Murray:2006ip,Ross:2010ia}. Thus accurate identification of patients with high risk of developing MS is crucial to limiting the disease activity and prolonging a `normal' patient lifestyle.

Early MS diagnosis is a difficult problem as it lacks a single diagnostic test and common clinical features are shared with other diseases. Neurologists rely primarily on either the Poser or McDonald diagnostic criteria to classify the disease. The Poser criteria separates MS into four groups based on attacks, clinical evidence, and paraclinical evidence~\cite{Poser:1983uw}. The McDonald diagnostic criteria, developed in 2001, leverages advancements in magnetic resonance imagining (MRI) techniques to facilitate diagnosis in typical clinical presentations~\cite{McDonald:2001tk}. Recent modifications to McDonald criteria improve the classification applicability to pediatric, Asian, and Latin American communities~\cite{Polman:2011co}. Nonetheless, a neurologist still relies on performing an exclusion diagnosis in conjunction with the patient's symptoms and medical history.

The advent of electronic medical records (EMRs) has increased the availability of medical data. Consequently, data mining and machine learning techniques have been used to develop clinical decision support systems to aid medical professionals. The problem of identifying patients with high risk of MS is a prime candidate for using EMRs to develop a data-driven prediction model. This paper investigates the feasibility and performance of a predictive disease model based on existing EMRs. Although our work is limited to patient data over a 7-year period, we establish a sparse baseline risk prediction model and demonstrate reasonable classification accuracy. 

\section{Background and Related Work}
\label{sec:prior-work}

The exact nature and cause of MS is still unknown. Epidemiology studies have focused on discovering the variables that influence the development of MS. Prior research has identified genetic, environmental, and comorbidity risk factors that affect the disease incidence rates. These variables have been used to build models to predict the diagnosis and progression of the disease.

\subsection{Risk factors}
Genetic susceptibility to MS has been supported by the following risk factors: race, gender, and family history. Genetic epidemiology studies have demonstrated a rise in disease risk when a family member is affected with MS~\cite{Kahana:2000vu, Sadovnick:2002us, Compston:2008js, Ramagopalan:2011co}. The increase in risk is correlated with the degree of kinship~\cite{Kahana:2000vu, Sadovnick:2002us, Compston:2008js, Ramagopalan:2011co}. Furthermore, the familial implications may also pertain to other autoimmune diseases~\cite{Compston:2008js, Lauer:2010ep, Ramagopalan:2011co}. An individual's race, which is genetically determined, also factors into the development of the disease. Certain races, such as Asian, Native American, and African American, are less susceptible~\cite{Kahana:2000vu,Ramagopalan:2011co}. Additionally, MS predominantly afflicts females and exhibit an early onset of the disease than their male counterparts~\cite{Tintore:2009bd, Ramagopalan:2011co}.

The place of residence during a patient's formative years is one of the environmental factors in the development of MS. Studies have shown that immigrants migrating before adolescence acquire the risk associated with their new residence while immigrants moving after adolescence retain their risk of their original residence~\cite{Hutter:1996en, Ascherio:2007jz, Ramagopalan:2011co}. The effect of latitude and hygiene hypothesis may account for the geographic variations beyond genetic factors.

The latitude gradient, a rise in incidence and prevalance rates with an increase in latitude, was previously a prominent MS feature but has declined in recent years. Sunlight duration, sunlight intensity, and vitamin D levels have been proposed as potential explanations for this phenomenon~\cite{Hutter:1996en, Wingerchuk:2001ve, Ascherio:2007cd, Lauer:2010ep, Ramagopalan:2011co}. The seasonal vulnerability also demonstrates the importance of sun exposure as a patient is most vulnerable during the winter~\cite{Hutter:1996en}. However, there are notable exceptions to the latitude theory in costal regions of Norway and Japan where a high consumption of fish dampens the lack of sunlight exposure~\cite{Hutter:1996en, Ramagopalan:2011co}.

The hygiene hypothesis postulated that early exposure to various infectious agents protects the patient against risk of MS~\cite{Ascherio:2007cd, Lauer:2010ep, Ramagopalan:2011co}. One infection in particular, Epstein Barr Virus (EBV), has been heavily associated with MS. Individuals with high anti-EBV antibodies have an increased risk of MS~\cite{Ascherio:2010hl}. Additionally, contracting EBV at a later age also increases the likelihood of developing MS~\cite{Ascherio:2007cd, Lauer:2010ep, Ramagopalan:2011co}. Other strains of viruses and infections,  such as human herpesvirus 6 (HHV6) and \textit{Chlamydia pneumoniae}, have been proposed but lack sufficient evidence to support a casual effect on disease risk~\cite{Wingerchuk:2001ve, Lauer:2010ep}.

One consequence of the hygiene hypothesis is the relationship between vaccinations and the susceptibility to MS. Countries with higher hygiene standards generally mandate vaccine immunizations to reduce the number of infections. However, during the late 1990s, concerns grew over the hepatitis B vaccine increasing the risk for MS~\cite{Marshall:1998jw}. Although subsequent studies~\cite{Ascherio:2001bd, Confavreux:2001ff} failed to find a significant correlation between the vaccine and the development of MS, the hypothesis that vaccinations may influence the development of the disease should not be dismissed~\cite{Ascherio:2007cd}.

An individual's lifestyle, through diet and smoking habits, also factors into the disease risk. Individuals who consume non-marine meat have higher risk of developing the disease~\cite{Lauer:2010ep}. However, fish and seafood consumption protects against MS~\cite{Hutter:1996en, Ascherio:2007cd, Lauer:2010ep, Ramagopalan:2011co}. Marine life has a higher concentration of polyunsaturated fatty acids and antioxidants, which has anti-inflammatory properties that help suppress the disease process~\cite{Hutter:1996en, Ascherio:2007cd}. High consumption of saturated fatty acids during one's childhood may cause adolescent obesity, which is associated with an increased risk of MS~\cite{Ramagopalan:2011co}. Additionally, cigarette smoking has been shown to influence the development of MS~\cite{Ascherio:2007cd, Lauer:2010ep, Ramagopalan:2011co}. 

Shifts in an individual's hormone levels have also been suggested as factors in the disease process. A decrease in the number of MS relapses during pregnancy suggests the transient benefits of higher levels of estrogen~\cite{Ascherio:2007cd, Tintore:2009bd}. A study on British women showed that the recent use of oral contraceptives reduced the risk of MS~\cite{Alonso:2005ce}. However, a subsequent US study~\cite{Hernan:2000ul} was unable to obtain evidence that supported the benefits of oral contraceptives.

Other autoimmune disorders and specific cancers have been proposed as potential comorbidities to MS. In a paper that summarized the environmental features researched in etiological research on MS~\cite{Lauer:2010ep}, Lauer noted that inflammatory bowel disease (IBD), ulcerative colitis, and Type 1 diabetes have the strongest correlations to MS amongst the various autoimmune disorders. The paper also referenced potential associations with Hodgkin's, oral, and colon cancers with the caveat that there was insufficient evidence to support these connections.

\subsection{Predictive models}
Predictive studies have primarily focused on the progression of the disease. Bergamaschi et. al~\cite{Bergamaschi:2001ud} identified clinical features that could help predict the onset of secondary progression, defined by an increase in the Kurtzke's Expanded Disability Status Scale (EDSS), using patient data collected in the first year of the disease. The factors discovered in the study were then used to propose a Bayesian Risk Estimate for Multiple Sclerosis (BREMS) score to predict the risk of reaching the secondary progression~\cite{Bergamaschi:2006jw}. A recent study suggested the use of EDSS ranking to identify patients at risk for high progression rates 5 years from the onset of the disease~\cite{Hughes:2012fu}.

Scoring systems have also been developed to assess the risk of disability. A study showed that MS Functional Composite, originally proposed as a clinical outcome measure, could be used to determine risk of severe physical disability~\cite{Rudick:2001wn}. The Magnetic Resonance Disease Severity Scale (MRDSS) combined MRI measures into a composite score to predict the progression of physical disabilities~\cite{Bakshi:2008fx}. Bazelier~\cite{Bazelier:2012cc} derived a score using Cox proportional hazard models to estimate the long-term risk of osteoporotic and hip fractures in MS patients. Another study conducted by Margaritella et. al~\cite{Margaritella:2012ba} used Evoked Potentials score to predict the progression of disability and identify patients with benign MS. 

Limited research has been done with regards to predicting the risk of developing MS. One work predicted MS in patients with mono symptomatic optic neuritis using MRI examination findings, oligoclonal bands in cerebrospinal fluid (CSF), immunoglobulin (Ig) G index,  and the seasonal time of onset~\cite{Jin:2003hb}. Thrower~\cite{Thrower:2007js} suggested the use of clinical characteristics of optic neuritis and traverse myelitis to identify high-risk MS patients. More recently, De Jager et. al~\cite{DeJager:2009el} proposed a weighted genetic risk score (wGRS) based on genetic susceptibility loci in the context of environmental risk factors. However, prior research relies on specialized measurements that are performed to confirm a MS diagnosis. The approaches suggested do not generalize to all patients and fail to allow for early diagnosis and intervention of high-risk MS patients.

\begin{table*}[tb]
\caption{The list of MS features and their associated categories. The order of feature introduction is denoted by the number next to the category name.}
\subtable{
\begin{tabular}{l}
\hline
\multicolumn{1}{c}{Demographics (1) } \\
\hline
Gender \\
Ethnicity \\
Race \\
Age \\
\hline
\multicolumn{1}{c}{Family History (1) } \\ 
\hline
MS \\
Mental Illness \\
Colon Cancer \\
Breast Cancer \\
Lupus \\
Thyroiditis \\
Diabetes \\
Inflammatory Bowel \\
\hline
\end{tabular}
}
\subtable{
\begin{tabular}{l}
\hline
\multicolumn{1}{c}{Autoimmune (2) } \\ 
\hline
Inflammatory Bowel \\
Celiac \\
Uveitis \\
Thyroiditis \\
Lupus \\
Rheumatoid arthritis \\
Sjoren's syndrome \\
Bell's palsy \\
Guillain Barre \\
Diabetes \\
Vitamin D deficiency \\
\hline
\end{tabular}
}
\subtable{
\begin{tabular}{l}
\hline
\multicolumn{1}{c}{Microbial (3)} \\ 
\hline
Measles, mumps, rubella\\
Epstein Barr Virus \\
\hline
\end{tabular}
}
\subtable{
\begin{tabular}{l}
\hline
\multicolumn{1}{c}{Mental Illness (4)} \\ 
\hline
Bipolar \\
Schizophrenia \\
\hline
\end{tabular}
}
\subtable{
\begin{tabular}{l}
\hline
\multicolumn{1}{c}{Cancer (5)} \\ 
\hline
Lymphoma \\
Oral \\
Breast \\
Colon \\
\hline
\end{tabular}
}
\subtable {
\begin{tabular}{l}
\hline
\multicolumn{1}{c}{Vaccine (6)} \\ 
\hline
Hepatitis (A+B) \\
Diphteria, tetanus, pertussis \\
Polio \\
Influenza \\
Measles, mumps, and rubella \\
Varicella (chicken pox) \\
Meningococcal \\
Pneumococcal \\
\textit{Haemophilus influenzae} type b\\
Human Papillomavirus \\
\hline
\end{tabular}
}
\subtable{
\begin{tabular}{l}
\hline
\multicolumn{1}{c}{Reproductive (7)} \\ 
\hline
Hysterectomy \\
Oral contraceptive pills \\
Estrogen replacement therapy \\
\hline
\end{tabular}
}
\subtable{
\begin{tabular}{l}
\hline
\multicolumn{1}{c}{MRI Scans + Obesity (8)} \\ 
\hline
Obesity \\
Abnormal brain MRI \\
Brain MRI \\
Cervical spine MRI \\
Thoracic spine MRI \\
\hline
\end{tabular}
}
\subtable{
\begin{tabular}{l}
\hline
\multicolumn{1}{c}{Blood Tests (9)} \\ 
\hline
Erythrocyte sedimintation rate \\
Lyme \\
B12 \\
ANA panel \\
Anti-cardiolipin antibody \\
Zinc \\
Cerebrospinal fluid exam \\
\hline
\end{tabular}
}
\label{tab:variables} 
\end{table*}

\section{Materials and Methods}

\subsection{Data}
Our retrospective study used de-identified patient data from the NorthShore Enterprise Data Warehouse (EDW). The data was collected from January 2006 to July 2012 and contained information pertaining to demographics, medications, medical encounters and procedures.

The study examined adults ($\geq$ 18 years of age) with complete demographic data (age, gender, ethnicity, and race). Any individual diagnosed with an MS ICD-9 code (``340") during a Neurology office visit was selected as a case patient. 1,456 case patients were identified in the NorthShore EDW. However, only 737 of the patients had recorded electronic medical encounters prior to the initial diagnosis date. For each of the 737 case patients, four control subjects with matching age and gender were selected from the general population for a total of 3,685 patients.

\subsection{Predictor Variables}
A comprehensive list of potential features was curated from prior MS research, detailed in section \ref{sec:prior-work}. The list was also expanded to include common vaccinations, cancers, mental illnesses, and autoimmune diseases. Unfortunately some of the variables, such as lifestyle factors (smoking and alcohol use) and diet were unrecorded in the EMR. In addition, some diseases were also excluded because none of the patients received the particular medical diagnosis during the study period. Table \ref{tab:variables} enumerates the features used in our retrospective study.

\begin{figure}[]
\begin{center}
\includegraphics[scale=0.40]{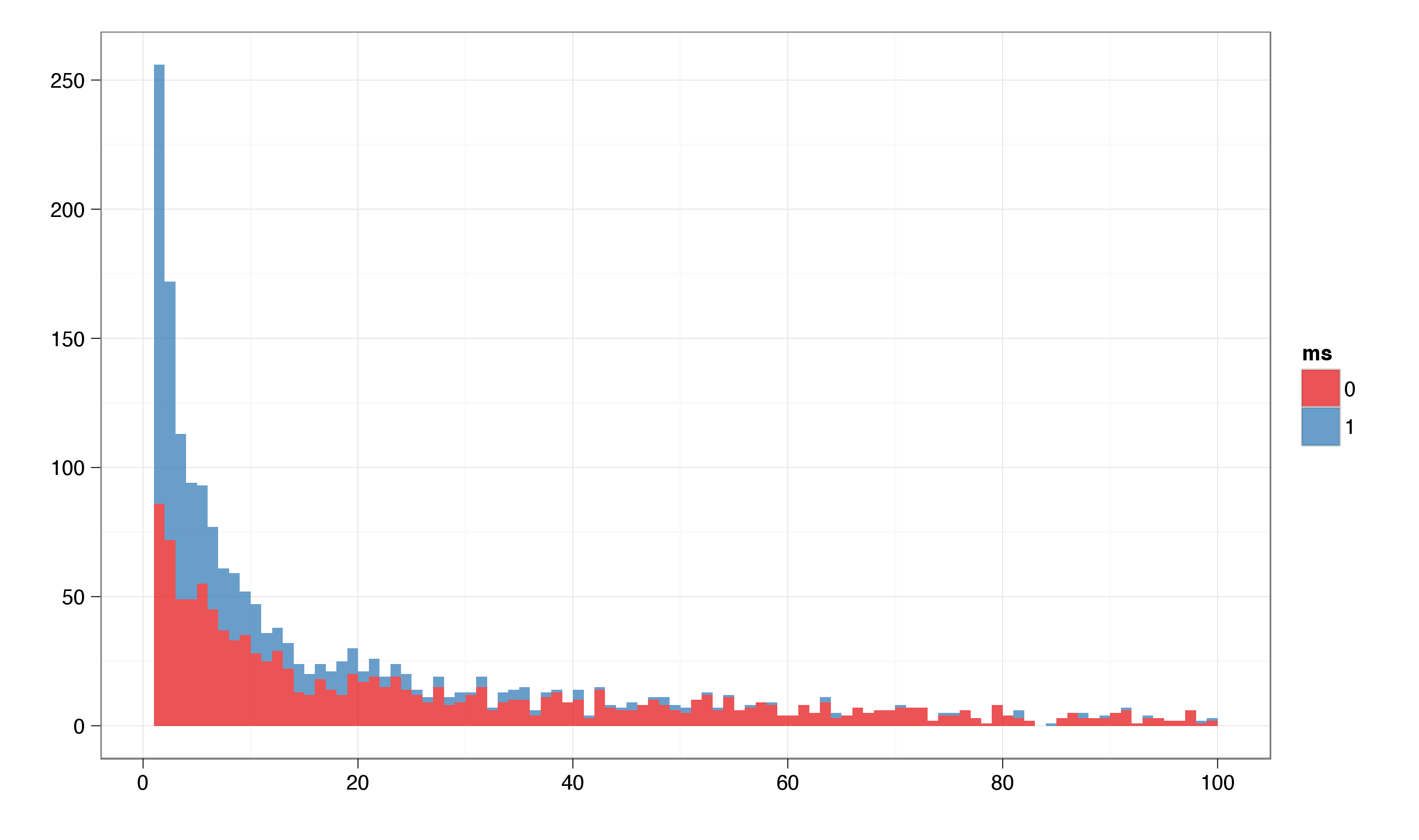}
\end{center}
\caption{Histogram of the number of encounters before $t_0$}
\label{fig:encounter}
\end{figure}

The initial MS diagnosis date is used to define $t_0$ for case patients. Since control patients did not have a MS diagnosis, $t_0$ is a randomly selected from the patient's encounter date. Figure \ref{fig:encounter} shows the frequency of the number of encounters with the same date and location prior to $t_0$. Case patients generally have a low number of previous encounters while there is a more even distribution amongst the control patients. An alternative option was to use the same $t_0$ for matching control patients. However, this exacerbated the discrepancy in the number of encounters prior to $t_0$ between case and control. Thus, a random encounter date was used to define $t_0$ for control patients.

All data, except those related to family history, obtained after $t_0$ were discarded to limit the potential effects of confounding factors. Family medical history spanned the entire study period because collection time is unimportant. A patient may fail to disclose all the family history in the first few medical encounters but reveals the information in later encounters after being diagnosed with a certain disease.

Binary values were used to indicate the presence of a particular medical diagnosis (ICD-9 code) found in a patient's encounter data prior to $t_0$. The vaccination, reproductive, and family medical history was extracted in a similar fashion, denoting the existing of specific supplemental classification codes (ICD-9 V codes). MRI scan features, except for an abnormal MRI result which was extracted via an ICD-9 code, signified the presence of specific medical procedure requests. Given the sparsity of numeric blood tests results, the feature was converted to three levels: (1) Unobserved, (2) Observed-Normal, and (3) Observed-Abnormal based on the ranges provided by the EDW. The entire feature set was represented using a binary matrix, where categorical variables were converted to dummy variables.

\subsection{Model Development}
Multivariate logistic regression models were used to predict a MS diagnosis at the next encounter. The selection of logistic regression model was motivated by the popularity of the model in the medical community, the simplicity of the model, and the interpretability of the results. To evaluate the effect on accuracy of specific predictor categories, new features were introduced in the order defined by Table \ref{tab:variables}. The first feature set contained only demographic and family history data, and was designed to mimic the information available at the first office visit. The last feature set contains all the predictor variables listed in Table \ref{tab:variables}. For each feature set, two sets of logistic regression models were trained: (1) forward stepwise model selection by Akaike information criterion (AIC) and (2) backward stepwise model selection by AIC. Stepwise model selection by AIC is used to minimize the model complexity, or encourage a sparser feature representation, without sacrificing predictive performance. 10-fold cross validation was used to estimate the accuracy of each model.

\section{Results}

\subsection{Feature Set Comparisons}
\begin{figure}[]
\begin{center}
\includegraphics[width=0.49\textwidth]{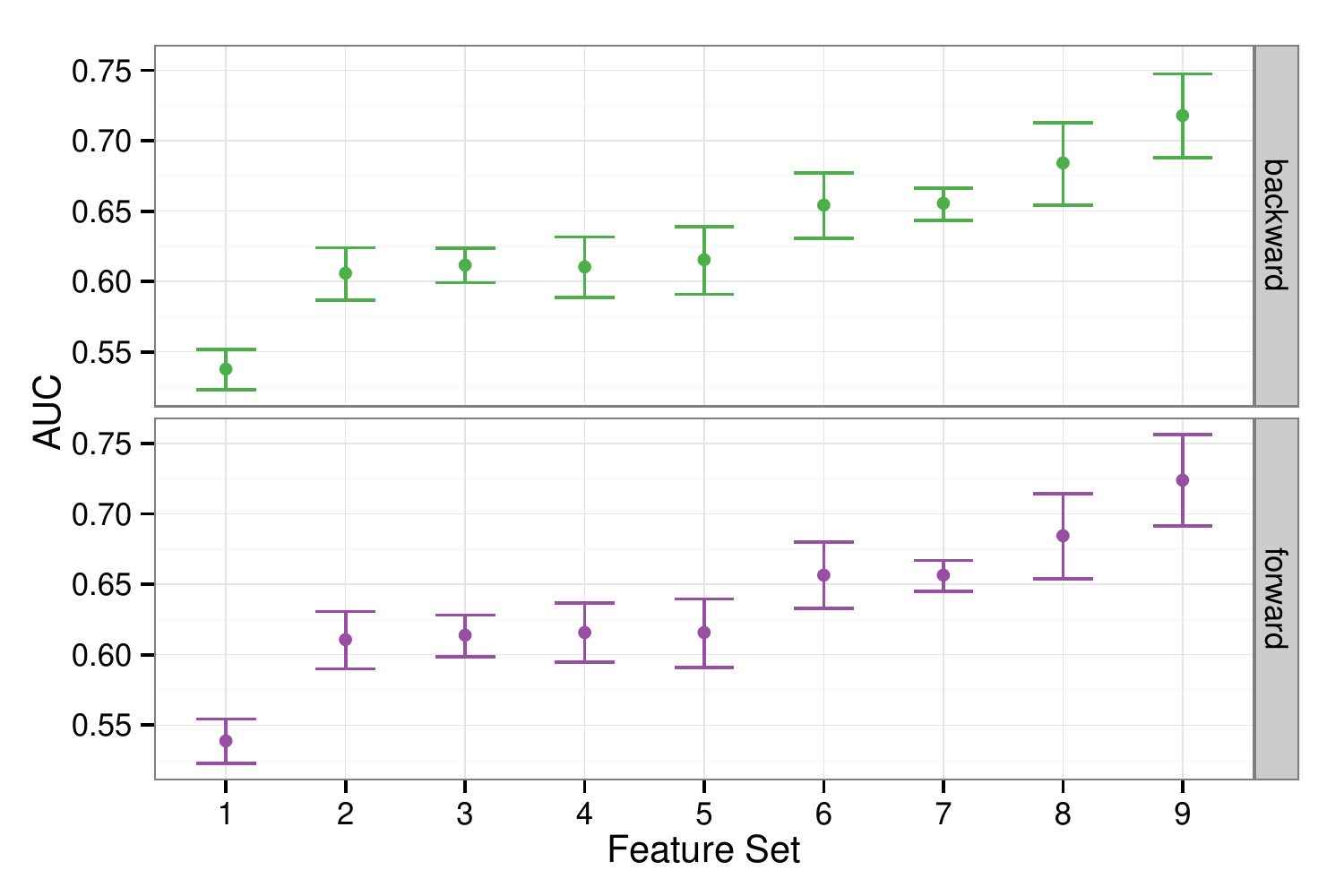}
\end{center}
\caption{A comparison of the AUC as categories of variables are added to the feature set.}
\label{fig:auc}
\end{figure}

Figure~\ref{fig:auc} shows the area under the receiver operating characteristic curve (AUC) for each feature set and model selection. Both feature selection methods result in similar predictive performance. Given only data that is available at the first office visit (demographic and family history), the forward selection model with an AUC of $0.538 \pm 0.016$ marginally outperforms random guess. Feature set 2, an expansion of the features to include autoimmune disorder diagnoses, increases the AUC by $0.072$. The performance then remains stagnant with the addition of the microbial, mental illness,  and cancer feature categories. However, vaccinations, MRI scans, and blood test results boosted the predictive performance. Using all the available features (feature set 9), the forward and backward feature selection models predict an MS diagnosis at the next visit with an AUC of $0.724 \pm 0.033$ and $0.718 \pm 0.030$ respectively.

\begin{figure}[]
\begin{center}
\includegraphics[width=0.49\textwidth]{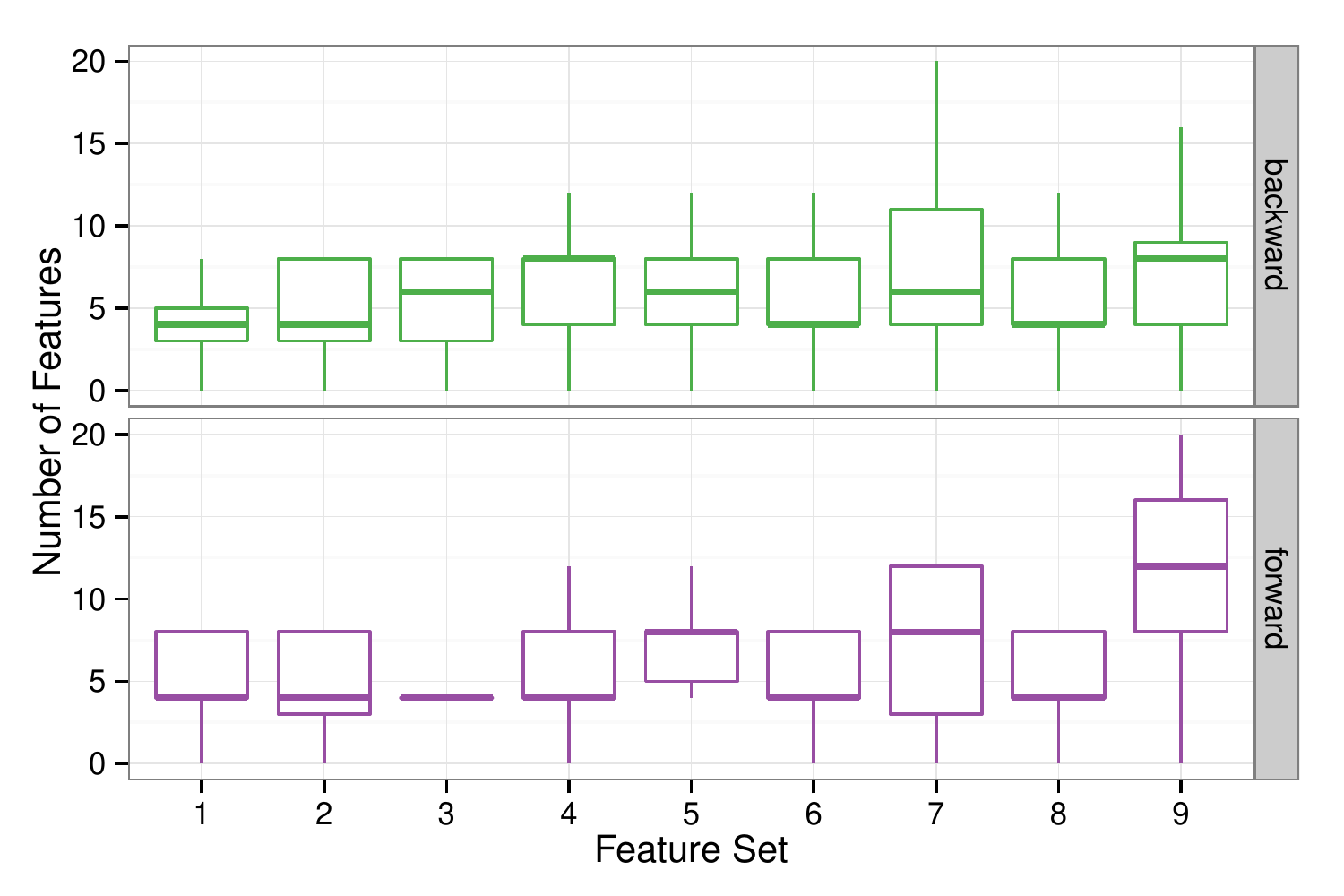}
\end{center}
\caption{A box plot of the number of variables selected per feature set and selection method.}
\label{fig:features}
\end{figure}

Stepwise feature selection using AIC produces a model using a sparse set of features. Figure \ref{fig:features} displays the comparison for the number of selected variables per feature set. The number of features selected with backward stepwise regression remains fairly constant for each feature set. This suggests the later categories (blood test results and MRI scans) are more informative. In addition, for feature sets 7-9, the backward selection method results in a sparse set of features. Both selection models on feature set 9 select less than 15\% of the potential features.

\begin{figure*}[htb]
\begin{center}
\includegraphics[width=0.99\textwidth]{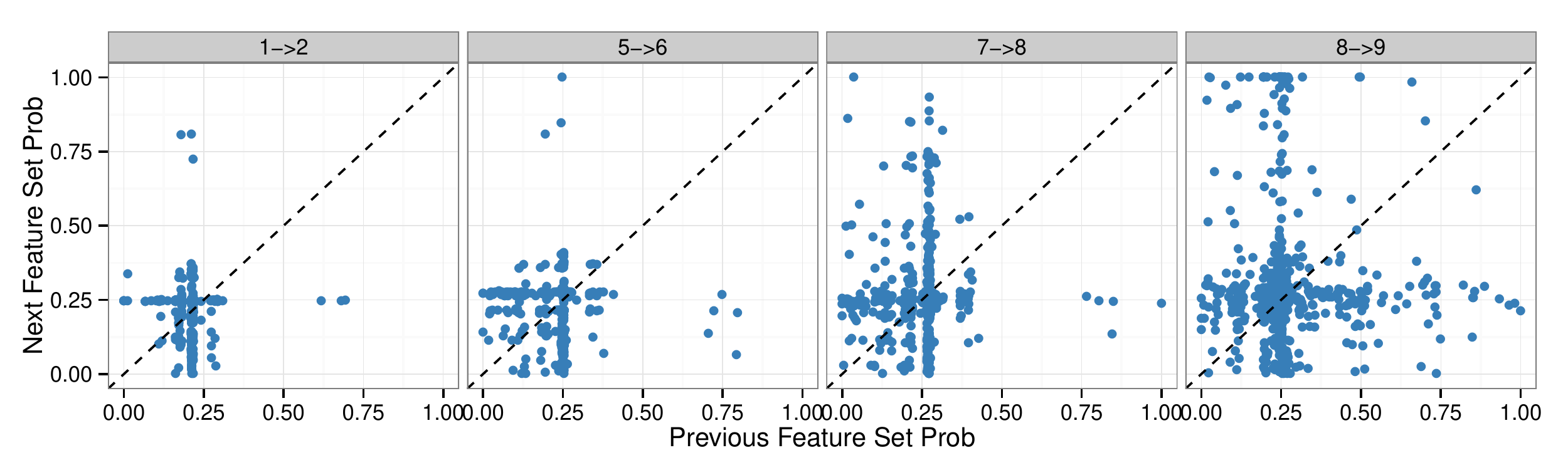}
\end{center}
\caption{A scatterplot comparison of the predicted probabilities between consecutive feature sets for 737 case patients. The dotted line signifies no change in predicted risk.}
\label{fig:positive-fs}
\end{figure*}

\begin{figure*}[htb]
\begin{center}
\includegraphics[width=0.99\textwidth]{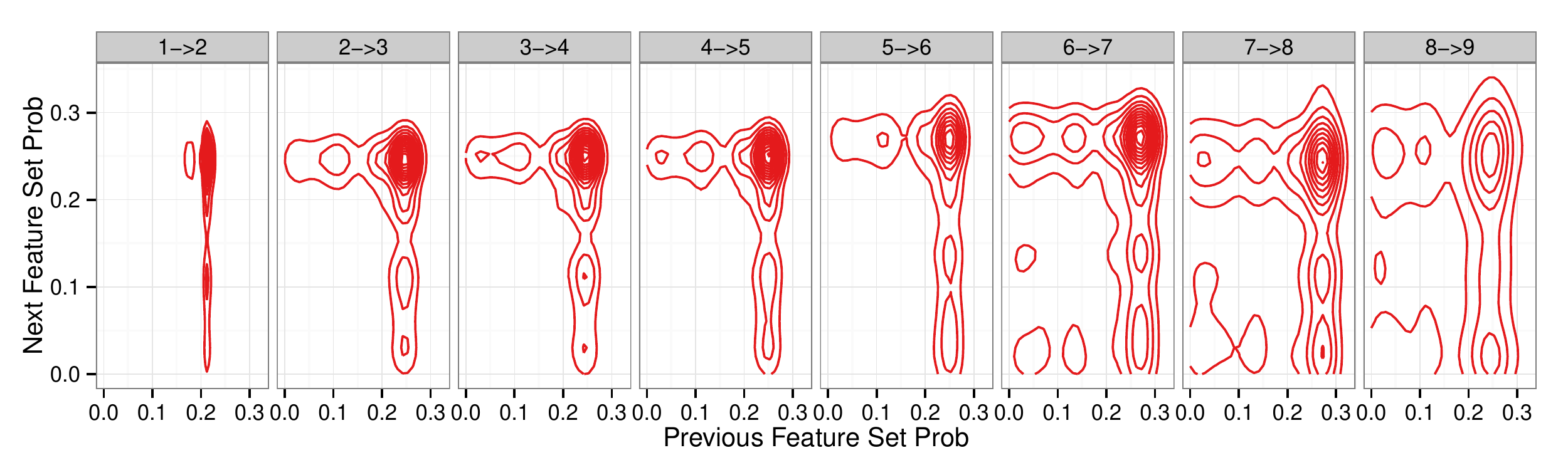}
\end{center}
\caption{A 2D kernel density estimate of the previous feature set and next feature set predicted probabilities for 2,948 control patients.}
\label{fig:negative-fs}
\end{figure*}

Figure \ref{fig:positive-fs} compares the joint predicted probabilities of consecutive feature sets for case patients and illustrates the effects of adding specific feature categories. Some of the transitions have been omitted since they are similar to the first feature set transition (1$\rightarrow$2). The addition of autoimmune disease diagnoses (1$\rightarrow$2) generally increases the predicted risk. The trend is most noticeable in the transition plot from feature 7 to 8, where the points lie predominately above the dotted line. Inclusion of blood test results (8$\rightarrow$9) marginally improves the predicted risk but it also decreases a substantial portion of patients with high risk in the previous model.

A joint density estimate of two consecutive feature sets for the control patients is demonstrated in Figure \ref{fig:negative-fs}. For the first transition (1$\rightarrow$2), the improvement to predictive performance can also be traced to the decrease in the predicted risk of low risk MS patients. In addition, the figure illustrates that as the feature is expanded, the density slowly shifts away from the top right corner to the bottom half of the plot. The transition from feature set 8 to 9 shows that the predicted risk is distributed more evenly for the control patients compared to the first feature set, where the probabilities are lie around 0.21.

Figures \ref{fig:positive-fs} and \ref{fig:negative-fs} show the dispersion of risk probabilities as more features are introduced. The addition of features related to MRI scans, or feature set 8, improves the separation between case and control through higher predicted probabilities for high-risk patients. The AUC improvement obtained from the inclusion of blood test results, feature set 9, can be primarily attributed to the decrease in predicted risk of control patients. These figures provide a graphical analysis of the benefits of specific feature categories.

\subsection{Feature Set 9 Results}
We further focus on the model with the best predictive performance, the forward selection model trained on all the features (feature set 9). Table \ref{tab:coeff} summarizes the variables for which the magnitude of the coefficient is larger than 1 in the majority of the folds. The table also displays the number of case and control patients with the feature, the odds ratio, and the p-value from a chi-square test to determine the significance of the variable. If we use p-value as an initial filter with $\alpha$=0.05, the following features would be eliminated: EBV; Bell's Palsy; colon cancer; family history of mental illness, MS, and inflammatory bowel disease; varicella vaccine, schizophrenia, and the \textit{Haemophilus influenzae} type b vaccine. However, the selection of some of these variables, such as history of mental illness and varicella vaccine, is surprising given the lack of sufficient evidence in prior work to support their effect on the development of MS.

\begin{table*}[htb]
\begin{center}
\begin{tabular}{l r r r r r}
\hline
Feature & \multicolumn{1}{c}{Beta}& \multicolumn{1}{c}{Case} 	&  \multicolumn{1}{c}{Control} &  \multicolumn{1}{c}{Odds ratio} &  \multicolumn{1}{c}{p-value}\\
\hline
Presence CSF oligoclonal bands	&16.255$\pm$0.545 	& 27		& 0		& $\infty$	& 0.000 \\
Mental Illness (FH)				& 6.298$\pm$3.101 		& 3		& 1		& 3		& 0.033 \\
EBV							& 3.974$\pm$3.924		& 3		& 2		& 1.5		& 0.093 \\
Abnormal brain MRI				& 2.877$\pm$0.313 		& 10		& 4		& 2.5		& 0.000 \\
Unobserved B12 				& 2.527$\pm$0.149 		& 674	& 2619	& 0.257	& 0.047 \\
Obs-Normal B12 	 			& 2.375$\pm$0.175		& 54		& 162	& 0.333	& 0.071 \\
Obs-Normal ANA SSB 			& 1.141$\pm$0.269 		& 57		& 95		& 0.6		& 0.000 \\
Bell's Palsy					& 1.889$\pm$0.295 		& 2		& 3		& 0.667	& 0.576 \\
Diabetes						& -1.036$\pm$0.078		& 19		& 224	& 0.085	& 0.000 \\
Obs-Normal ANA DS			&-1.066$\pm$0.122		& 44		& 78		& 0.564	& 0.000 \\
Oral contraceptive				&-1.244$\pm$0.205		& 2		& 35		& 0.057	& 0.043 \\
DTP vaccine 					& -1.829$\pm$0.126		& 12		& 327	& 0.037	& 0.000 \\
Unobserved Lyme Test 			& -1.980$\pm$0.099 	& 692	& 2924	& 0.237	& 0.000 \\
Colon cancer					&-2.927$\pm$4.072 		& 2		& 15		& 0.133	& 0.584 \\
Asian race					&-2.925$\pm$0.234 		& 2		& 93		& 0.022	& 0.000 \\
MS (FH)						&-3.356$\pm$4.307 		& 2		& 19		& 0.105	& 0.023 \\
Unobserved CSF IGG synthesis	&-4.841$\pm$0.296		& 700	& 2946	& 0.238	& 0.000 \\
Varicella vaccine				&-15.161$\pm$0.087 	& 0		& 13		& 0		& 0.145 \\
HPV vaccine					&-15.728$\pm$2.188 	& 0		& 82		& 0		& 0.000 \\
Schizophrenia					&-15.763$\pm$0.369	& 0		& 10 		& 0 		& 0.235 \\
Estrogen replacement			&-15.823$\pm$0.209	& 0		& 22		& 0 		& 0.037 \\
IBD (FH)						&-17.236$\pm$0.420	& 0 		& 3 		& 0 		& 0.885 \\
HIB vaccine					&-18.156$\pm$0.521	& 0 		& 2 		& 0 		& 1.000 \\ 
\hline
\end{tabular}
\end{center}
\caption{Mean coefficient values, odds ratio, and p-value for variables picked in at least 5 of the 10 folds with $|\beta| > 1$}
\label{tab:coeff}
\end{table*}

\begin{figure}[htb]
\begin{center}
\includegraphics[width=0.49\textwidth]{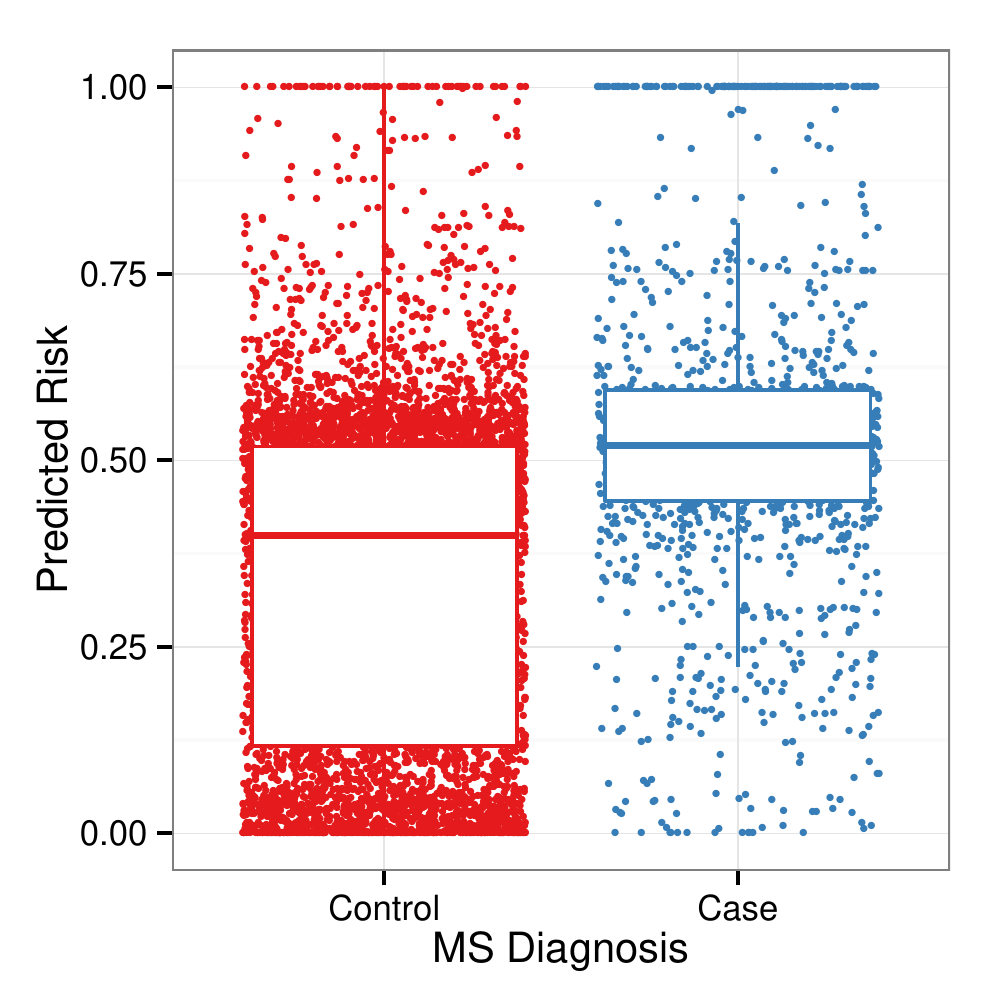}
\end{center}
\caption{Box plot of the predicted probabilities using a forward selection model on all the features.}
\label{fig:risk}
\end{figure}

Figure \ref{fig:risk} shows the distribution of predicted risk values. The figure shows that even with all the features listed in Table \ref{tab:variables}, there is still a considerable overlap of predicted probabilities between case and control patients. Better separation between these two classes can improve the risk prediction accuracy. The plot suggests incorporation of additional diagnoses or temporal aspects of existing diagnoses may be necessary to improve model performance.

\begin{figure*}[htb]
\subfigure[ROC curves compared to random assignment]{
\includegraphics[width=0.48\textwidth]{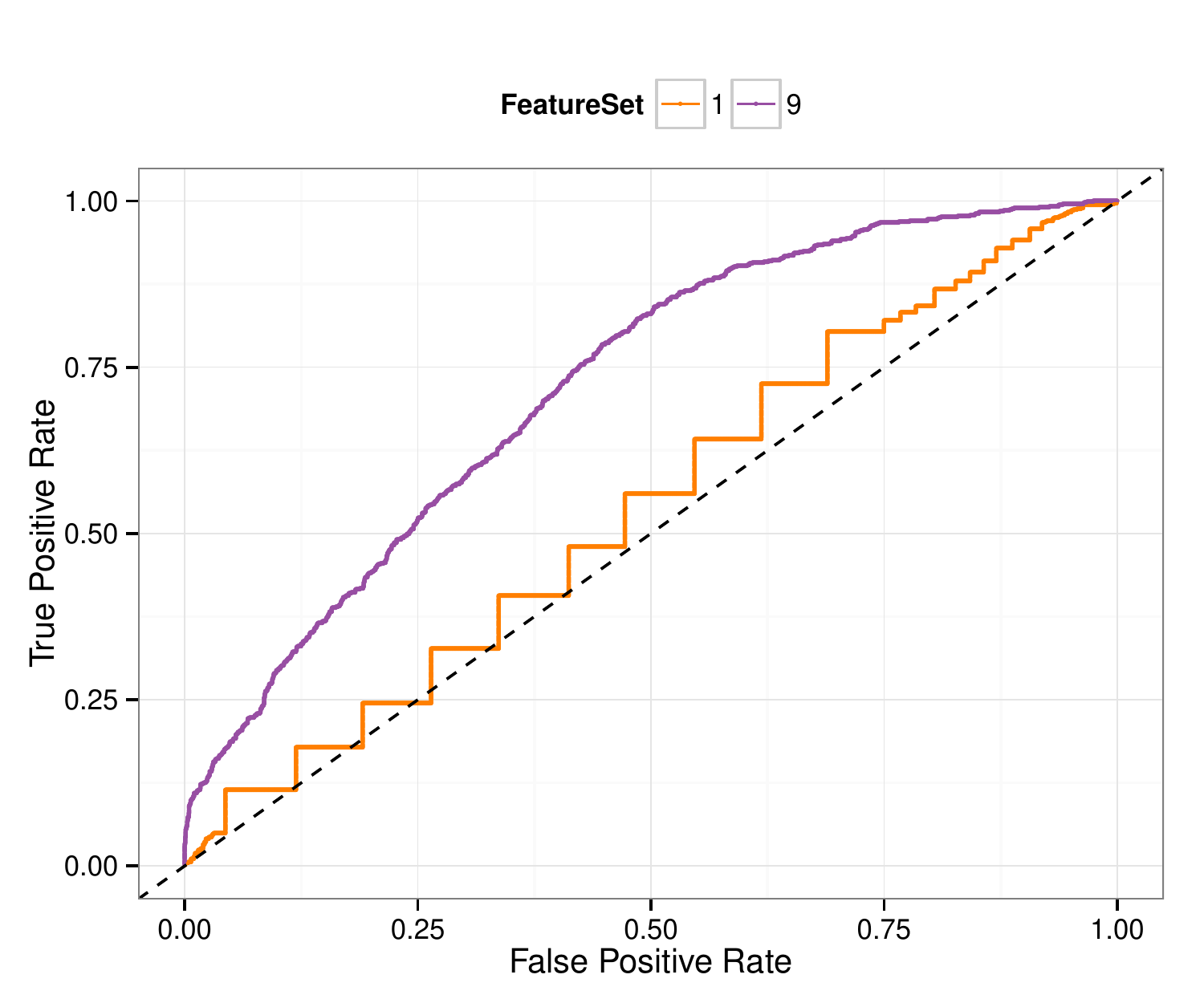}
\label{fig:roc}
}
\subfigure[Sensitivity, specificity, and positive predictive value as function of threshold]{
\includegraphics[width=0.48\textwidth]{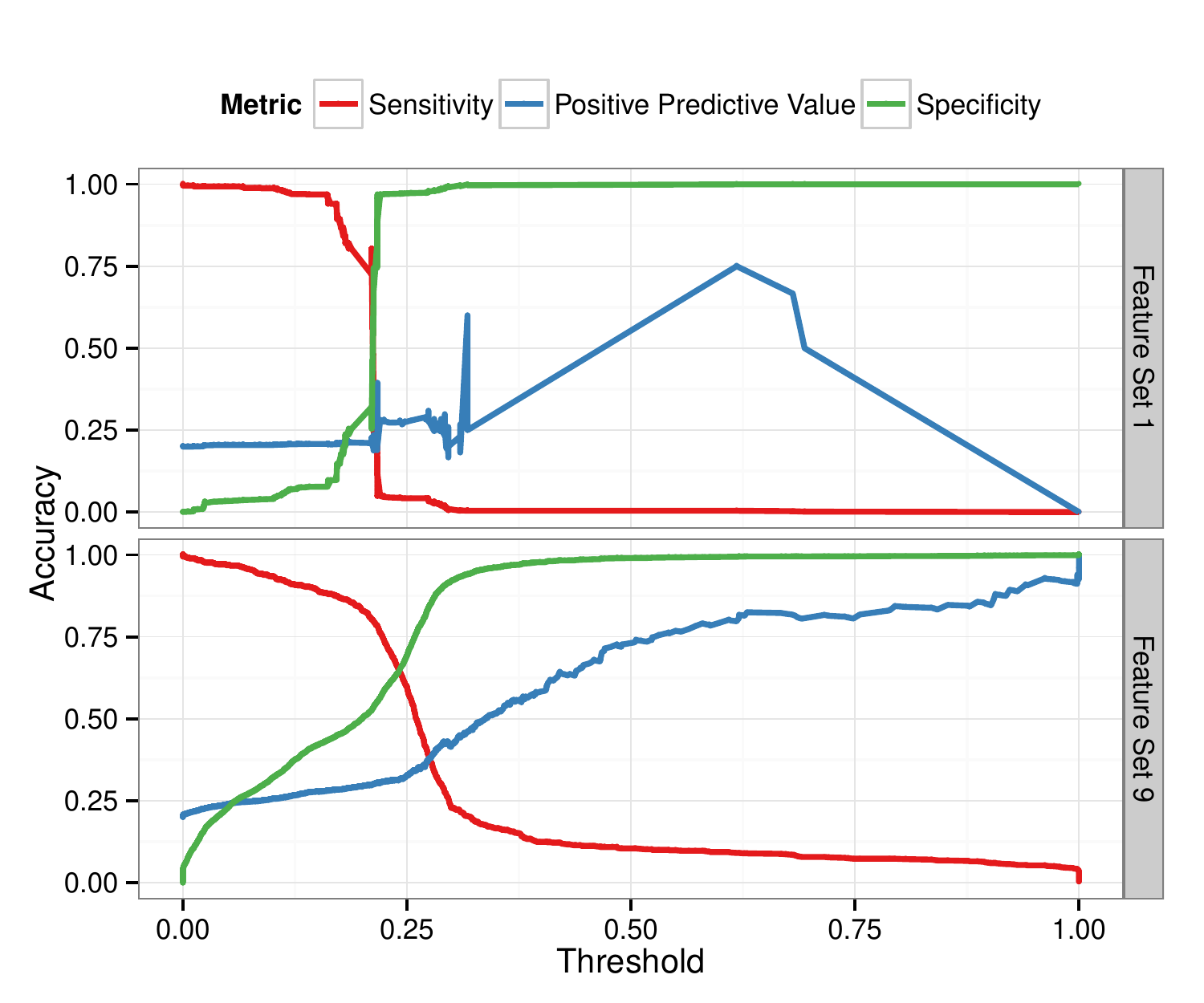}
\label{fig:ppv-sens}
}
\caption{Model performance plots for feature sets 1 and 9.}
\label{fig:perf}
\end{figure*}

Figure \ref{fig:perf} contains the performance plots for the forward selection models trained on feature set 1 and feature set 9. Figure \ref{fig:roc} demonstrates the noticeable improvement using all the available features. Additionally, the model trained on feature set 1, demographics and family history features, barely outperforms random chance. The tradeoff between sensitivity, specificity, and positive predictive value can be seen in Figure \ref{fig:ppv-sens}. Feature set 9 has a higher intersection between the sensitivity and specificity curves, which is summarized in Table \ref{tab:intersect}. In addition, the full-featured model generally achieves a better positive predictive value for all threshold values. However, the positive predictive value and sensitivity curves cross at the value $\sim 0.40$. At this point, we can accurately diagnose 40\% of the case patients, but only 2 out of every 5 patients predicted to have a high risk of MS will be diagnosed with MS at the next office visit, a high number of false positives.

\begin{table}[htb]
\begin{center}
\begin{tabular}{l c c c c}
\hline
Feature Set & Cutoff & Sensitivity & Sensitivity & PPV \\
\hline
1 & 0.212 & 0.528 & 0.528 & 0.218\\
9 & 0.241 & 0.647 & 0.647 & 0.314 \\
\hline
\end{tabular}
\end{center}
\caption{The intersection of the sensitivity and specificity curve from Figure \ref{fig:ppv-sens}.}
\label{tab:intersect}
\end{table}
\subsection{Discussion}
The results demonstrate reasonable predictive accuracy using all the available features. One potential hindrance lies in the current feature construction. As Figure \ref{fig:encounter} shows, there are a limited number of encounters prior to $t_0$ for case patients. Thus, it is difficult to determine whether an unobserved diagnosis may be due to the lack of longitudinal data (the patient was diagnosed prior to the study period). Additionally, certain diagnoses, such as EBV, can only be verified through culture samples which are not performed for every patient.

Another limitation of our study is the reliance on ICD-9 and procedure codes. A patient may exhibit all the clinical symptoms for a specific disease but it is not present in the encounter data because the disorder has not been diagnosed. The ambiguity of ICD9-codes and diagnostic discrepancies between medical doctors can also impact our feature construction. Moreover, the blood test results' conversion to a categorical feature may be inaccurate as the testing protocol may have changed during the study window. Therefore, a patient's feature vector may not accurately reflect their medical history.

Our study also suggests incorporating additional features. Given that some of the variables were unrecorded in the structured portion of the EMR, parsing through the clinical notes could result in information regarding lifestyle factors, diet, detailed family and medical history. In addition, temporal aspects of the medical diagnoses were not included in our feature set since the data was confined to medical encounters over a 6-year period. 

\section{Conclusion}
This paper presented a risk prediction model from EMRs to help address the difficulty of early diagnosis in MS patients. A sparse set of features were selected to minimize model complexity while maintaining reasonable predictive performance. Our results show we are able to help identify patients at high-risk of developing MS, in spite of a limited sample of patient data. In addition, our models have the ability to generalize to other healthcare systems as we rely only on components commonly found in electronic patient data.

The work demonstrates the potential of leveraging EMRs to aid medical professionals with difficult tasks, especially with early disease diagnosis. Future work will focus on incorporating temporal components, such as time of diagnosis, into the model, decreasing the false positive rate, and integrating a larger control population.

\section{Acknowledgments}
We thank Afif Hentati and Demetrius "Jim" Maraganore for their guidance, advice and comments on this study. We acknowledge comments from and conversations with Kibaek Kim, Yubin Park, Xiang Zhong, and Sanjay Mehrotra. We are indebted to Justin Lakeman for extracting data from the NorthShore Enterprise Data Warehouse.

\bibliographystyle{IEEEtran}
\bibliography{IEEEabrv,MS12}
\end{document}